\documentclass[superscriptaddress,twocolumn]{revtex4-1}

\usepackage{epsfig}
\usepackage{amsmath}
\usepackage{graphicx}
\usepackage{dcolumn}
\usepackage{amsmath}
\usepackage{latexsym}
\usepackage{epsfig}
\usepackage{amsmath}
\usepackage{graphicx}
\usepackage{dcolumn}
\usepackage{amsmath}
\usepackage{latexsym}
\usepackage{color}
\usepackage{epstopdf}
\usepackage{subfigure}
\usepackage{comment}
\usepackage{nicefrac}
\usepackage{epstopdf}
\usepackage{citesort}

\usepackage[ulem=normalem]{changes}
\usepackage{float}

\usepackage{xcolor}
\usepackage{soul}

\begin{document}

\title{The Higgs Mode in Disordered Superconductors\\ Close to a Quantum Phase Transition}

\author{Daniel Sherman \dag}
\affiliation{Department of Physics, Bar Ilan University, Ramat Gan 52900, Israel}
\affiliation{1.~Physikalisches Institut, Universit{\"a}t Stuttgart, Pfaffenwaldring 57, 70550 Stuttgart, Germany}
\author{Uwe S. Pracht}
\affiliation{1.~Physikalisches Institut, Universit{\"a}t Stuttgart, Pfaffenwaldring 57, 70550 Stuttgart, Germany}
\author{Boris Gorshunov}
\affiliation{1.~Physikalisches Institut, Universit{\"a}t Stuttgart, Pfaffenwaldring 57, 70550 Stuttgart, Germany}
\affiliation{Moscow Institute of Physics and Technology, 141700, Dolgoprudny, Moscow Region, Russia}
\affiliation{Prokhorov Institute of General Physics, Russian Academy of Sciences, Vavilov Street 38, 119991 Moscow, Russia}
\author{Shachaf Poran}
\affiliation{Department of Physics, Bar Ilan University, Ramat Gan 52900, Israel}
\author{John Jesudasan}
\affiliation{Tata Institute of Fundamental Research, Homi Bhabha Road, Colaba, Mumbai 400005, India}
\author{Madhavi Chand}
\affiliation{Tata Institute of Fundamental Research, Homi Bhabha Road, Colaba, Mumbai 400005, India}
\author{Pratap Raychaudhuri }
\affiliation{Tata Institute of Fundamental Research, Homi Bhabha Road, Colaba, Mumbai 400005, India}
\author{Mason Swanson}
\affiliation{Department of Physics, The Ohio State University, Columbus, OH 43210, USA}
\author{Nandini Trivedi}
\affiliation{Department of Physics, The Ohio State University, Columbus, OH 43210, USA}
\author{Assa Auerbach}
\affiliation{Physics Department, Technion, 32000 Haifa, Israel}
\author{Marc Scheffler}
\affiliation{1.~Physikalisches Institut, Universit{\"a}t Stuttgart, Pfaffenwaldring 57, 70550 Stuttgart, Germany}
\author{Aviad Frydman}
\affiliation{Department of Physics, Bar Ilan University, Ramat Gan 52900, Israel}
\author{Martin Dressel}
\affiliation{1.~Physikalisches Institut, Universit{\"a}t Stuttgart, Pfaffenwaldring 57, 70550 Stuttgart, Germany}

\maketitle

\textbf{The concept of mass-generation via the Higgs mechanism was strongly inspired by earlier works on the Meissner-Ochsenfeld effect in superconductors. In quantum field theory, the excitations of longitudinal components of the Higgs field manifest as massive Higgs bosons. The analogous Higgs mode in superconductors has not yet been observed due to its rapid decay into particle-hole pairs. Following recent theories, however, the Higgs mode should decrease below the pairing gap $2\Delta$ and become visible in two-dimensional systems close to the superconductor-insulator transition (SIT).  For experimental verification, we measured the complex terahertz transmission and tunneling density of states (DOS) of various thin films of superconducting NbN and InO close to criticality. Comparing both techniques reveals a growing discrepancy between the finite $2\Delta$ and the threshold energy for electromagnetic absorption which vanishes critically towards the SIT. We identify the excess absorption below $2\Delta$ as a strong evidence of the Higgs mode in two dimensional quantum critical superconductors.} \\

\begin{figure*}
\begin{center}
\includegraphics[scale=0.08]{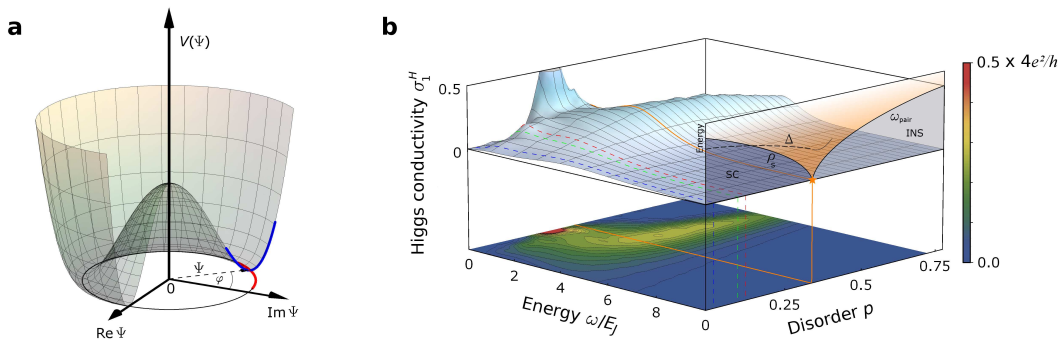}
\caption{\label{fig:HiggsCond}(Color online) \textbf{Broken $U(1)$-symmetry phase and quantum Monte Carlo calculation of the Higgs conductivity}.\textbf{a.} When symmetry is broken, the potential acquires a \emph{Mexican hat} shape with a circle of potential minima along the brim (black solid circle). Transverse modes of the order parameter $\Psi=Ae^{i\varphi}$ along the brim (red line) are Nambu-Goldstone (phase) modes, and longitudinal modes (blue line) are Higgs (amplitude) modes associated with a finite energy. In superconductivity, the potential corresponds to the free energy. \textbf{b.} The Higgs mode gives rise to low-frequency conductivity (in units of $4e^2/h$), that grows as disorder $p$ (fraction of disconnected superconducting islands) is increased and remains finite through the quantum phase transition (orange line). At the quantum critical point, $p_c=0.337$, the superfluid density, $\rho_s$, in the superconducting phase vanishes and the quasiparticle gap, $\Delta$, remains finite whereas in the insulator $\omega_\mathrm{pair}$, which is the energy to insert a Cooper-pair to the insulator, goes to zero. Results for specific disorder (blue, green, and red dashed lines) and compared to experiment, see Fig. \ref{spectral}. For details of the calculation see \cite{Swanson}. }
\end{center}
\end{figure*}

The Higgs mechanism, which has great implications to recent developments in particle physics \cite{Review}, originates in Anderson's pioneering work on  symmetry breaking with gauge fields in superconductors \cite{anderson}. A superconductor spontaneously breaks continuous $U(1)$ symmetry and acquires the well-known \emph{Mexican hat} potential with a degenerate circle of minima described by the order parameter $\Psi=Ae^{i\varphi}$, see Fig. \ref{fig:HiggsCond}a. Excitations from the ground state can be classified as transverse Nambu-Goldstone (phase) modes and massive longitudinal Higgs (amplitude) modes (see blue and red lines in Fig. \ref{fig:HiggsCond}a). In particle physics, the latter manifest themselves as the Higgs boson which was recently discovered at CERN \cite{Higgs-CERN}. Indications of a Higgs mode in correlated many-body systems have been found in one-dimensional charge-density-wave systems \cite{CDW}, quantum antiferromagnets \cite{Ruegg}  and two-dimensional superfluid to Mott transition  in cold atoms \cite{higgs-coldatmos}. An amplitude mode, also named \emph{Higgs mode}, was theoretically predicted for superconductors \cite{VarmaLittlewood} and recently reported to be measured by pump-probe spectroscopy \cite{shimano}. This amplitude mode describes pairing  fluctuations, which are qualitatively distinct from the purely bosonic  mode expected  from the O(2) field theory. The Higgs-amplitude mode analogous to the high-energy Higgs Boson has not yet  been  observed in superconductors. A partial reason is that in homogeneous, BCS superconductors the Higgs mode is short-lived and decays to particle hole (Bogoliubov) pairs \cite{Sachdev, Zwerger}. Nevertheless, collective modes were recently predicted to be significant in strongly disordered superconductors \cite{lara}, and, in particular it was shown \cite{podolsky,dynamics, gazit} that the Higgs mode softens but remains sufficiently sharp near a quantum critical point (QCP) in two dimensions since it is found to be a critical energy scale of the quantum phase transition. Hence, the Higgs mass can be reduced below twice the pairing gap, $2\Delta$, making this  mode experimentally visible. Such a critical point has been suggested to be relevant for  the SIT in two dimensional films.

The desired quantum phase transition (QPT) from a superconductor to an insulator can be tuned by introducing disorder on atomic length scales. It has been shown both experimentally \cite{kowal,sacepe,Kamlapure,sherman-tunneling,madhavi} and theoretically \cite{ghosal1,ghosal2,dubi} that though being morphologically homogeneous, with increasing disorder superconducting films can progressively become electronically granular on length scales comparable to the superconducting coherence length. While for modest disorder the superconducting state is hardly affected, strong disorder near the QCP decomposes the homogeneous state into individual superconducting islands. In this scenario, the QPT takes place at the critical disorder when phase fluctuations between different islands destroy the global phase coherence and the superfluid density $\rho_s$ vanishes on a macroscopic length scale \cite{Nandini}. Consequently, the loss of global phase coherence does not necessarily cause the {\em pairing} \emph{gap} $\Delta$ to close as the decoupled islands still remain superconducting. The value of the critical temperature $T_c$ in the vicinity of the QPT is thus not defined by the opening of a gap in the quasiparticle density of states, but rather by the presence of a global phase coherence. Indeed, finite values of $\Delta$ in strongly disordered thin films were experimentally observed in tunneling spectroscopy experiments where $T_c$ was already vanishingly small on the superconducting side or even zero on the insulating side of the QPT \cite{pratap,sherman-tunneling}. Near the QPT one expects  two critical energy scales:  on the insulating side, a charge gap $\omega_\mathrm{pair}$, which is the energy required to insert a Cooper-pair into the pair insulator \cite{Nandini}, and on the superconducting side, the Higgs (amplitude) mass gap. Both energy scales should vanish at the QPT.\\

\begin{figure*}
\vspace{0cm}
\includegraphics[scale=0.57]{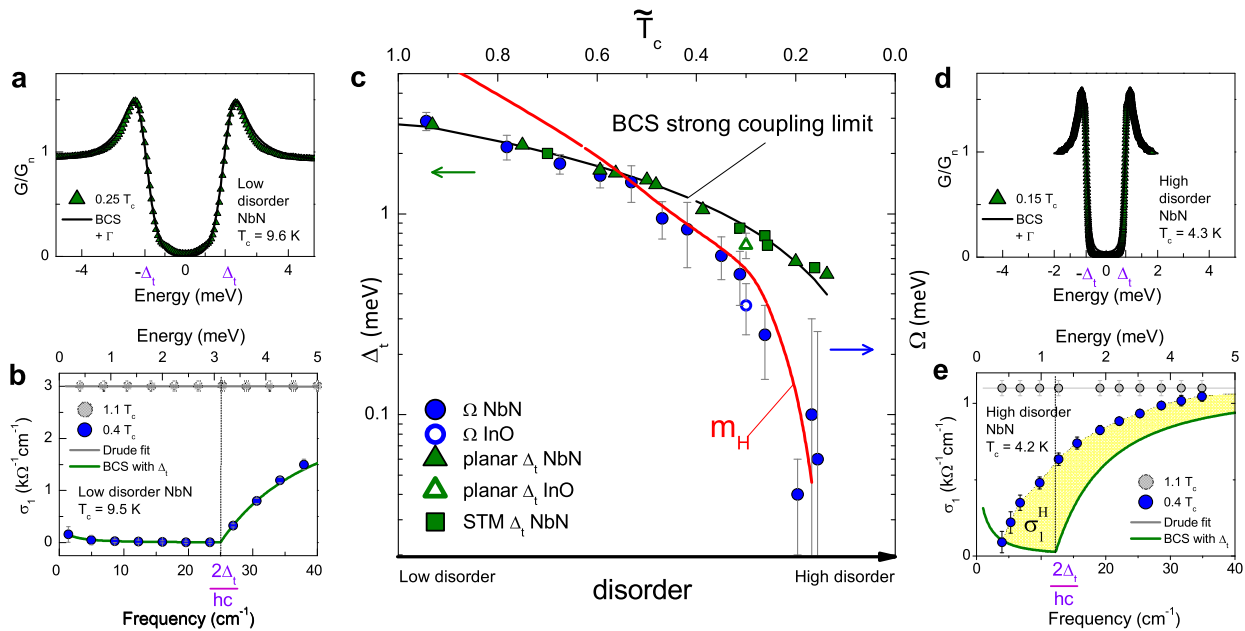}
\vspace{-3.5cm}
\caption{\textbf{Tunneling versus optical spectroscopy. a,b} Experimental results on low disordered NbN samples. Panel (a) shows the measured tunneling conductance spectrum (green triangles) alongside a fit to BCS (black line) with a Dynes broadening parameter, $\Gamma$. Panel (b) shows the real part of the dynamical conductivity, $\sigma_1$, versus frequency (energy) at temperatures below and above $T_c=9.5K$. The low temperature curve is fitted (green line) to Mattis-Bardeen while plugging in the energy gap value obtained in the corresponding tunneling result, $\Delta_t$. \textbf{c}, summary of the quasiparticle tunneling gap, $\Delta_t$ (green symbols), measured by planner tunneling junctions or scanning tunneling microscope (STM), versus $\Omega$, the frequency at which $\sigma_1(\omega)$ is minimal (blue symbols) obtained from optical spectroscopy for several superconducting NbN and InO films spanning the different degrees of disorder. While the quasiparticle gap, $\Delta_t$, remains fairly unchanged with increasing disorder and basically falls on the BCS strong coupling limit ratio, $\Omega$ is significantly suppressed. According to Mattis-Bardeen theory, for ideal superconductors $\sigma_1$ is minimal at a frequency $\Omega$ that corresponds to $2\Delta$. The discrepancy between both spectroscopic probes increases towards the highly disordered limit signaling the presence of additional modes superposing the quasiparticle response. The solid red line corresponds to the analytical prediction of $m_H$ close to a QPT calculated by Podolsky et. al. \cite{podolsky}.\textbf{d,e,} Experimental results on high disordered NbN samples. Panel (d) shows the measured tunneling conductance spectrum (green triangles) alongside a fit to BCS (black line) with a Dynes broadening parameter, $\Gamma$. Panel (e) shows the real part of the dynamical conductivity, $\sigma_1$, versus frequency (energy) at temperatures below and above $T_C=4.2K$. The low temperature curve is fitted (green line) to Mattis-Bardeen while plugging in the energy gap value obtained in the corresponding tunneling result. Unlike in the case of the low disordered sample, these two curves differ. The excess spectral weight, marked in yellow and defined as the difference between the curves, is attributed to the Higgs contribution, $\sigma_1^H$, see text.
The error bars for $\sigma_1$ in the graphs are determined by the distortion of the Fabry Perot oscillations due to parasatic radiation, standing waves and electronic noise}
\label{fig:tun_thz}
\end{figure*}

Assuming the presence of a Higgs mode  in the superconducting thin film, what would be the most suited experimental quantity to detect it? The Higgs mode is a finite-energy oscillation of the order parameter magnitude $|\Psi|$. It can be probed by the dynamical conductivity $\hat{\sigma} (\omega)$ which depends on the current-current correlation function $\langle [j(t),j(0)] \rangle$. At low temperatures, the current is dominated by the Cooper pair current $j\sim(2e)\mathrm{Im}\{\Psi^*\nabla\Psi\}\simeq(2e)|\Psi|^2\nabla\varphi$, where $\varphi$ is the local phase field and e the elementary charge. As a result, the conductivity  depends on a convolution of the amplitude and phase fluctuations. \\

How would the Higgs mode contribute to the dynamical conductivity? Theoretically it is predicted to give rise to excess conductivity at sub-gap frequencies \cite{podolsky} which we will refer to as \emph{Higgs conductivity}, $\hat{\sigma}^H(\omega)$, in the remainder of the paper. In non-disordered systems \cite{dynamics} $\sigma_1^H(\omega)$ shows a hard gap at frequencies similar to the superconducting gap, $\omega \sim 2\Delta/\hbar$, that is associated with the energy scale of the Higgs mode, $m_H$. This gap becomes softer as the system approaches the QPT, reaching zero at the critical point.  Recently, Swanson and collaborators \cite{Swanson} studied the effect of disorder on the dynamical conductivity across the superconductor-insulator QPT employing quantum Monte Carlo methods and extracted the excess low-frequency contribution, see Fig. \ref{fig:HiggsCond}b. The calculations show that the presence of disorder suppresses $m_H$  so that $\sigma_1^H(\omega)$ remains finite across the QPT. This excess conductivity adds to the conductivity stemming from the superfluid condensate and the quasiparticle dynamics, so that one can write
\begin{equation}
\hat{\sigma}(\omega)=\sigma_1(\omega)+i\sigma_2(\omega)= \underbrace{A \rho_s \delta(\omega)+\hat{\sigma}^{qp}(\omega)}_{\hat{\sigma}^{\mathrm{BCS}}(\omega)}+\hat{\sigma}^H(\omega) \label{eq:1}
\end{equation}
where $\rho_s$ is the superfluid density and $A$ is a constant,\cite{DresselBook}.   \\

In order to experimentally search for the contribution of the Higgs mode, we have studied disordered superconducting films of NbN and InO by means of THz spectroscopy. Since the superconducting energy gaps are of the order of $0.1-1$ THz, optical spectroscopy in this regime is an alternative method to tunneling spectroscopy for the measurement of  $2\Delta$. Most important, unlike tunneling which measures the density of states of the quasiparticles, optical spectroscopy probes a complex response function, $\hat{\sigma}^\mathrm{exp}$, that constitutes from the superfluid condensate, the quasiparticle dynamics and collective modes, see Eq. (\ref{eq:1}). One can decompose the optically measured conductivity into the regular BCS contribution and the contribution of the collective excitations. The first one is modeled by the Mattis-Bardeen (MB) theory for ordinary superconductors using our tunneling spectroscopy results as input to fix the absolute numbers. The difference to the experimental data determines the Higgs mode simply by calculating
\begin{equation}
\sigma^H_1(\omega)=\sigma^\mathrm{exp}_1(\omega)-\sigma^{\mathrm{BCS}}_1(\omega). \label{eq:2}
\end{equation}

\begin{figure*}
\vspace{-3cm}
\includegraphics[scale=0.55]{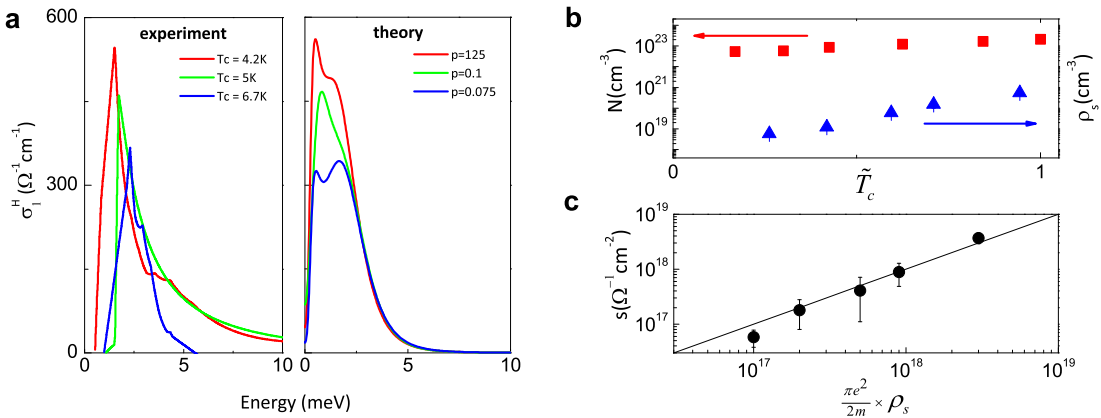}
\vspace{-3cm}
\caption{\textbf{Higgs conductivity and spectral weight. }\textbf{a}, Experimental and theoretical results  for the Higgs conductivity $\sigma_1^H$ as a function of energy for three NbN films of different disorder. The numerical results \cite{Swanson} were obtained for a fixed value of $E_C/E_J$, whereas the degree of disorder, reflecting breaking bonds between the superconducting islands, is denoted by $p$. Qualitative and quantitative features are shared by both experiment and theory. The sharp lines in the experiment data are due to interpolation between measured data points. \textbf{b}, Charge carrier density $N$ in the normal state obtained from Hall measurements (black squares) and superfluid density, $\rho_s$, measured by optical spectroscopy as functions of $T_c/T_c^{\mathrm{clean}}$ (reflecting the degree of disorder).  Note the faster decrease of $\rho_s$ with increasing disorder, indicating the vanishing contribution of the superfluid condensate to the spectral weight. \textbf{c}, the redistribution of the 'missing' spectral weight $s$ between normal and superconducting state versus the superfluid density $\rho_s$, as defined in Eq. (\ref{eq:SW}). The observed linear relation indicates that the redistribution of the spectral weight occurs within our measured energy spectrum.}
\label{spectral}
\end{figure*}

We have measured the complex transmission coefficient of several thin-film samples with different degrees of disorder utilizing Mach-Zehnder interferometry. Measurements were performed in the frequency domain between $0.05-1.2$\,THz (corresponding to $1.7 - 40$  cm$^{-1}$
or  $0.18 - 5\,$ meV) for temperatures above and well below $T_c$. From this we directly obtain the real and imaginary parts, $\sigma_1^\mathrm{exp}$ and $\sigma_2^\mathrm{exp}$, of the dynamical conductivity, in a individual manner without Kramers-Kronig analysis. According to MB theory, $\sigma_1$ is minimal at a frequency $\Omega$ that corresponds to twice the superconducting energy gap, $2\Delta$. Furthermore, the superfluid density is related to $\sigma_2(\omega)$
\begin{equation}
\rho_s=\frac{\sigma_2(\omega)m\omega}{e^2}\label{eq:SF},
\end{equation}
where $m$ is the electron mass and $e$ is the elementary charge. This robust approach is well established to study superconducting thin films. For more details see the methods section and, e.g., \cite{Uwe-IEEE,dressel2,DresselBook,Uwe-TiN}.
Fig. \ref{fig:tun_thz}b,e shows the real part of the conductivity $\sigma^\mathrm{exp}_1(\omega)$ for modestly ($T_c=9.5$\,K) and strongly ($T_c=4.2$\,K) disordered NbN in the normal state and well below $T_c$ together with the fits to the Mattis-Bardeen prediction for the disordered regime \cite{Zimmermann91}.
In both cases, $\sigma^\mathrm{exp}_1(\omega)$ is featureless in the normal state following a simple Drude behaviour with a scattering rate well above the THz range, whereas $\sigma^\mathrm{exp}_1(\omega)$ is strongly suppressed in the superconducting state. The ordered sample is fitted perfectly by the Mattis-Bardeen theory. The onset of the high-frequency upturn coincides with twice the energy gap, $\Delta_t$, obtained by tunneling spectroscopy performed on a similar sample \cite{pratap}, as seen in Fig.\ref{fig:tun_thz}a. The situation is remarkably different for the strongly disordered sample. Here the decrease towards low frequencies is not at all captured by BCS theory (green curve). In fact, using $\Delta_t$ extracted from corresponding tunneling experiment, as seen in Fig.\ref{fig:tun_thz}d, yields a curve which is significantly below $\sigma^\mathrm{exp}_1(\omega)$. With increasing disorder, both the discrepancy between $2\Delta_t$ and $\Omega$ and  the insufficiency of Mattis-Bardeen fits  become progressively worse. This trend is demonstrated in Fig. \ref{fig:tun_thz}c where we compare results from both techniques on a large number of NbN and InO samples spanning the various degrees of disorder (measured in terms of the normalized critical temperature, $\tilde{T}_c=T_c/T_c^{\mathrm{clean}}$). For small disorder, $\tilde{T}_c\simeq 1$, tunneling and THz spectroscopy yield the same value for the superconducting energy gap. Upon increasing disorder (decreasing $\tilde T_c$) the discrepancy becomes more and more pronounced. For the most-disordered samples, we find about one order of magnitude difference between corresponding values. We assign these differences to an absorption process stemming from the Higgs mode that becomes progressively prominent as the system approaches the quantum critical point. This explains the discrepancy in the sense that $\Omega$ in the strong-disorder limit no longer equals $2\Delta$ as a consequence of the additional conductivity $\sigma_1^H(\omega)$ of the emergent Higgs mode. The previously prominent spectral feature marking the gap frequency is now hidden in the shoulder at higher frequencies.  Although a distinct experimental determination of $\Omega$ becomes progressively  difficult as it is pushed to low frequencies,  we note the resemblance between $\Omega$ and the theoretical prediction of $m_H$  in the vicinity of the critical point \cite{podolsky}, as seen in Fig. \ref{fig:tun_thz}c.\\

We now explore the evolution of the observed additional excess weight associated with the Higgs conductivity, $\sigma^H_1(\omega)$, as defined in Eq. (\ref{eq:2}), and compare these measured results with recent numerical simulations detailed in ref. \cite{Swanson} and sketched in Fig. \ref{fig:HiggsCond}b. Fig. \ref{spectral}a shows the measured $\sigma_1^H(\omega)$ for three disordered NbN films with different $T_c=6.7,\,5$ and $4.2$\,K  and the theoretical calculation for corresponding values of disorder $p=0.075,\,0.1$ and $0.125$.  We note that one cannot expect a perfect quantitative agreement since the theory assumes that $2\Delta$ is much larger than the Higgs mode energy, whereas experimentally they are of the same order of magnitude. Nevertheless, the overall behaviour and even quantitative trends are shared by theory and experiment: There is a pronounced peak of $\sigma_H(\omega)$ that shifts towards smaller frequencies and becomes sharper with increasing disorder. \\
The appearance of the Higgs mode must go along with a redistribution of the spectral weight as this quantity is strictly conserved; it measures the total charge carrier density $N$ in the system \cite{DresselBook}. In accordance with the bosonic model of the SIT sketched above, the strength of the $\delta$-peak, i.e. the superfluid density $\rho_s$, dwindles to zero in the vicinity of the quantum critical point. Fig. \ref{spectral}b displays $\rho_s$ for disordered NbN films extracted from the imaginary part of the conductivity, using Eq. (\ref{eq:SF}) and $N$ in the normal state obtained from Hall measurements. While $\rho_s$ is reduced by about 2 orders or magnitude with increasing disorder, $N$ is much less affected. According to the Ferrell-Tinkham-Glover sum rule \cite{DresselBook} for the 'missing' spectral weight $s$ between normal and superconducting states,
\begin{equation}
s=\int_{0^+}^\infty \mathrm{d}\omega[\sigma^n_1(\omega)-\sigma^s_1(\omega)]\sim\rho_s, \label{eq:SW}
\end{equation}
a reduced superfluid density $\rho_s$ upon increasing disorder leads to a reduced value of $s$. As the quasiparticle gap remains fairly unchanged with disorder, this necessarily causes the spectral weight contribution of the Higgs mode, $\int \mathrm{d}\omega\sigma^H_1(\omega)$, to become more pronounced. Fig. \ref{spectral}c depicts the detected linear relation between the missing spectral weight, $s$, and the superfluid density, $\rho_s$, for several films with different degree of disorder, thus providing the self consistency of the above argument and eliminating the possibility of a redistributed spectral weight to higher frequencies (due to a sudden change in the scattering rate, for example).\\

We conclude that the low-frequency absorption observed by optical spectroscopy originates from the Higgs mode in superconductors close to a quantum phase transition. As the system approaches the critical point, the energy scale for this mode decreases and its magnitude grows, exhibiting quantitative agreement with numerical simulations.

The study of the properties of disordered superconductors is a subject of ongoing intense activity, mostly because it is viewed as being one of the few physical systems that can be tuned through a two dimensional quantum critical point, which is not mean-field-like. The softening of the Higgs mode is a  direct proof that the SIT transition is a quantum critical point  in which a diverging timescale is detected. Evidently, the vicinity to the QPT offers a unique opportunity to study the nature of the low energy collective excitations in superconductors. Going beyond disordered superconductors, our findings can play a role in tracing collective excitations in other quantum critical condensed matter systems and might influence related fields such as Bose-condensed ultracold atoms, quantum statistical mechanics and high energy physics.
\vspace{0.5cm}

\textbf{Methods}

The InO films were deposited on $10$x$ 10$\,mm$^2$ of THz-transparent MgO or sapphire substrates (with various thickness ranging from 0.5 to 1.5\,mm) by e-gun evaporation. During the deposition process dry oxygen was injected into the chamber; the partial oxygen pressure allows us to tune the disorder. The NbN films were grown on similar MgO substrates by reactive magnetron sputtering, where the Nb/N ratio in the plasma served as a disorder tuning parameter. In both cases the deposited films were structurally homogeneous; the thickness ranges from 15 to 40\,nm. DC transport measurements were used to characterize $T_c$.
THz spectroscopy was applied in the past to confirm the BCS theory since it probes the energy range of the superconducting gap \cite{Uwe-IEEE,dressel2,Uwe-TiN}. The experimental setup \cite{dressel2,Uwe-IEEE} is based on several backward wave oscillators as a powerful radiation sources to emit continuous-wave, coherent radiation which, in sum, can be tuned over the frequency range of $0.05 - 1.2$\,THz, corresponding a photon energy of $0.18 - 5$\,meV.
We employ a quasi-optical Mach-Zehnder interferometer to measure the complex transmission $T=te^{i\theta}$ with $t$ the amplitude and $\theta$ the phase shift of radiation passing through the sample under study from which the complex conductivity, $\hat{\sigma}(\omega)$, is directly calculated. The samples were mounted in an optical $^4$He cryostat with a continuously accessible temperature range spanning from 300 to 1.85\,K. To further proceed with the experimental data, we employ two analysis routines. In the first one, $t$ and $\theta$ are simultaneously fitted to a combination of Fresnel equations (for multiple reflections) \cite{DresselBook} for the optics and a suited microscopic model for the charge carrier dynamics (i.e. Drude theory for the metallic and BCS theory for the superconducting state complemented by a finite scattering rate \cite{Zimmermann91}). Free-electron parameters (such as the scattering rate or plasma frequency) required for the BCS fit are taken from Drude fits to the normal-state $t$ and $\theta$ slightly above the superconducting transition. The superconducting energy gap $2\Delta$ is then obtained as the sole fit parameter. While this approach is well established for BCS-type, i.e. non-disordered, superconducting systems, it fails for disordered systems beyond the Anderson limit. The second routine is suited for systems where no microscopic model is available, i.e. strongly disordered systems. In a narrow band around each Fabry-Perot resonance (which are caused by the finite thickness of the sample), we fit $t$ and $\theta$ to exclusively the Fresnel equations using $\sigma_1$ and $\sigma_2$ as fit parameters. Depending on the optical thickness of the substrate this routine yields 10 to 15 pairs of $\sigma_1$ and $\sigma_2$ for each resonance frequency $\omega_i$. Details of the experimental setup and analysis routines are found, e.g., in \cite{Uwe-IEEE,dressel2,DresselBook,Uwe-TiN,sherman-THz}

\vspace{0.5cm}
$\dag$ Present address: Center for Quantum Devices, Niels Bohr Institute, University of Copenhagen, 2100 Copenhagen, Denmark.

\vspace{1cm}

\textbf{Acknowledgements:} We are grateful for useful discussions with D. Arovas,  L. Benfatto, S. Gazit, D. Podolsky and E. Shimshoni.  We acknowledge support from the the GIF foundation grant I-1250-303.10/2014 and from the Deutsche Forschungsgemeinschaft. U.S.P acknowledges financial support from the Studienstiftung des deutschen Volkes. B.G. acknowledges support from the Russian Ministry of education and science (Program “5 top 100”), M.S acknowledges support from the NSF Graduate Research Fellowship, N.T. acknowledges support from grant DOE DE-FG02-07ER46423 (N. T.) and computational support from the Ohio Supercomputing Center and A.A. acknowledges support from the ISF and BSF foundations.

\vspace{1cm}

\textbf{Author Contributions} D.S., J.J., M.C. and P.R. carried out the DC experiments.  D.S., U.S.P. and B.G. carried out the THz experiments. D.S., U.S.P. and A.F. analyzed the data. D.S., S.P., J.J. and P.R. prepared the samples. N.T., M.S. and A.A. carried out the theoretical analysis and the numerical simulations. U.S.P., D.S., M.D., A.F., M.S., N.T. and A.A.  wrote the paper. A.F. and M.D. initiated and supervised the work. All the authors discussed the results and commented on the manuscript.

\vspace{1cm}

\textbf{Additional Information} The authors declare that they have no completing financial interests. Correspondence and
requests for materials should be addressed to A.F.

\end{document}